\documentclass[twocolumn,prb,superscriptaddress]{revtex4-2}
\usepackage{graphicx}
\usepackage{physics}
\usepackage{color}
\usepackage{amsmath}
\usepackage{amsfonts}
\usepackage{xspace}
\usepackage{array}
\usepackage{makecell}
\usepackage{ulem}
\usepackage[dvipsnames]{xcolor}
\usepackage[breaklinks,colorlinks,linkcolor=blue,citecolor=blue,urlcolor=blue]{hyperref}
\usepackage{ulem}
\def\avg(#1){\langle#1\rangle}

\def\Im{\rm{Im}}
\def\be{\begin{equation}}
\def\ee{\end{equation}}
\def\bea{\begin{eqnarray}}
\def\eea{\end{eqnarray}}

\def\nn{\nonumber}

\def\tr{\mbox{tr}}

\usepackage{ulem}

\begin{document}	
	
	\title{Dissipation induced Liouville-Majorana modes in open quantum system}%
	\author{Xing-Shuo Xu}
	\author{Xiang-Fa Zhou}
	\email{xfzhou@ustc.edu.cn}
	\author{Guang-Can Guo}
	\author{Zheng-Wei Zhou}
	\email{zwzhou@ustc.edu.cn}
	\affiliation{CAS Key Lab of Quantum Information, University of Science and Technology of China, Hefei, 230026, China}
	\affiliation{Synergetic Innovation Center of Quantum Information and Quantum Physics, University of Science and Technology of China, Hefei, 230026, China}
	\affiliation{ Hefei National Laboratory, University of Science and Technology of China, Hefei 230088, China}
	\date{\today}	
	
	\begin{abstract}
		
		In open systems, topological edge states quickly lose coherence and cannot be used in topological quantum computation and quantum memory. Here we show that for dissipative quantum spin (or fermionic) systems, topologically non-Hermitian Liouville-Majorana edge modes (LMEMs) can survive in the extended Liouville-Fock space, which is beyond the scope of topological modes defined in usual Hermitian system. By vectorizing the Lindblad equation of the system using the third quantization, we prove that it reduces to a series of non-Hermitian Kitaev chains in the extended Liouville-Fock space, and topologically LMEMs are protected due to its internal symmetry. Furthermore, we provide an explicit method for detecting  these modes and prove that the purity of the density matrix characterizes the long-range correlation of LMEMs. 	The work opens new avenues of searching for novel stable topological states in open systems induced by quantum jumps.
	\end{abstract}

	\maketitle
	
	\textit{Introduction.}
	The realization and manipulation of topological quantum states in various systems have received sustained attention in many different  fields of physics\cite{ozawa2019topological,luo2015quantum,bardyn2012majorana,stern2013topological,zhang2018topological,goldman2016topological,karzig2017scalable,hyart2013flux,alicea2011non,fraxanet2022topological}.
	Since topological phases possess nonlocal orders robust to local perturbations,  this intrinsic stability makes them ideal platforms for topological quantum computation and quantum memory.
	Meanwhile, the system's novelty also enables the construction of various quantum devices that traditional materials can not cover\cite{yan2012topological,culcer2020transport,bernevig2022progress}.
	On the other hand,  topological phases are inevitably coupled to their surroundings in natural systems.  The resulting quantum dissipation can destroy these phases and  spoil the signals induced by their topological features\cite{pichler2010nonequilibrium,yan2022experimental,poletti2013emergence,cai2013algebraic,syassen2008strong,sponselee2018dynamics,schmidt2012decoherence,tomita2017observation,sciolla2015two,tomita2017observation,henriet2019critical,seetharam2022correlation}.
	Therefore, searching for novel robust topological effects, even in  dissipation, becomes essential to implement various topological phases of matter and quantum computing tasks within current systems \cite{bouganne2020anomalous,diehl2011topology,bardyn2013topology,verstraete2009quantum}.  
	
	Topological physics in non-Hermitian dissipative systems has also been widely discussed recently\cite{shen2018topological,gong2018topological,song2019non,okuma2020topological,ghatak2019new,borgnia2020non,mi2022noise,maiellaro2022edge}. However, in most discussions, dissipation is characterized only by introducing an effective non-Hermitian Hamiltonian. The influence and back action of detections and quantum jumps on the system's dynamics  are only less considered. For a dissipative system under the Markovian approximation, the general dynamics are governed by Lindblad equations\cite{van2022finite,albert2014symmetries,chetrite2012quantum,prosen2008third,prosen2010spectral,prosen2010exact,daley2014quantum,de2021constructing,PhysRevB98094308,PhysRevB106L060307}, where both the dissipators and the influence of quantum jumps are explicitly considered. Although topological Majorana modes can  be stationary states of the system by carefully designing the dissipative Lindblad operators,  in general cases,  Majorana modes are unstable in the presence of dissipations\cite{kitaev2001unpaired,diehl2011topology,reslen2020uncoupled,PhysRevB98094308}. It is thus natural to ask: what topological properties will be stable in dissipative systems?
	Answering the question is a highly non-trivial task, as currently, solving the master equation for dissipative many-body systems is still a challenging task analytically and numerically\cite{carmele2015stretched,goldstein2019dissipation,huang2019dissipative}.
	Therefore, finding exactly solvable dissipative models with stable topological characteristics becomes a key ingredient in understanding non-trivial topological effects induced by dissipations, which is also less considered in current studies.
	
	In this work, we provide an analytically solvable model described by the Lindblad equation with site-dependent couplings and dissipations. Formally, this is achieved by vectorizing the density matrix, and mapping the Lindblad equation into a Schr\"{o}gdinger-like equation in the extended Liouville-Fock space  with effective non-Hermitian Hamiltonian\cite{prosen2008third,prosen2010spectral,prosen2010exact,reslen2020uncoupled}.  Therefore, topological properties discussed for non-Hermitian Hamiltonian can also be transplanted to open quantum systems described by Lindblad equations.
	The main results can be summarized as follows.
	\begin{enumerate}
		\vspace{-0.2cm}
		\item We prove the model maps to a series of non-Hermitian Kitaev chains in the extended Liouville-Fock space. Moreover, for open boundaries,  the system supports topological Liouville-Majorana edge modes (LMEM) beyond the scope of the usual Hermitian Majorara modes discussed in a closed system.
		\vspace{-0.2cm}
		\item The proposed LMEMs are robust to symmetry-preserving disturbances and can be verified by fixed ratios of physical observables under time evolution.  The correlations in LMEMs can also be distilled by quadratic forms of physical observables \cite{islam2015measuring,abanin2012measuring,cardy2011measuring,elben2018renyi,hastings2010measuring,rakovszky2019sub}. 
		\vspace{-0.2cm}
		\item Our work also highlights the importance of quantum jumps for implementing novel topological states in dissipative systems.
	\end{enumerate}

	\textit{The model and non-Hermitian Liouvillian.} We start by considering the Lindblad equation of the spin system subject to local dissipations
	\bea
	i\dot{\rho}=[H,\rho]+i\sum_{j=1}^N(L_{j}\rho L_{j}^{\dagger}-\frac{1}{2}\{L_{j}^{\dagger}L_{j},\rho\}),  \label{eq:master1}
	\eea
	where the Hamiltonian and Lindblad operators read
	\bea
	H=\sum_{j}^{N-1}J_{j}\sigma_{j}^{x}\sigma_{j+1}^{x}, \hspace{.25cm}
	L_{j}=\sqrt{\gamma_{j}}\sigma_{j}^{z}.
	\label{eq:model}
	\eea
	Here $J_j$ is the coupling strength between nearest-neighboring spins, and $\gamma_j$ is the local  dephasing rates. We note that in current system, all nontrivial dissipative dynamics is attributed to the presence of quantum jump terms $L_{j}\rho L_{j}^{\dagger}$, as the relevant non-Hermitian Hamiltonian contains only homogeneous dissipations due to $L^{\dag}_{j}L_{j}=L_{j}L^{\dag}_{j}=\gamma_j I_j$.
	
	Without dissipation, the model can be solved by introducing the celebrated Jordan-Wigner transformation as $\sigma_{j}^{x}=\prod_{k<j}(-iw_{2k-1}w_{2k})w_{2j-1}, \sigma_{j}^{y}=\prod_{k<j}(-iw_{2k-1}w_{2k})w_{2j}$.
	Here $w_{j}$ is the usual single-site Majorana fermion (MF) and satisfies $\{w_{i},w_{j}\}=2\delta_{ij}$.
	The Hamiltonian can be written as $H=\sum_{j}J_{j}iw_{2j}w_{2j+1}$, where two isolated edge MFs $w_{1}$ and $w_{2N}$ are decoupled with $H$ as $[H,w_{1}]=[H,w_{2N}]=0$,  and can be combined to form a Dirac fermion.
	Since $w_{1}$ and $w_{2N}$  are spatially separate, this fermionic excitation is nonlocal and robust to local perturbations, which can then be used as an ideal platform to encode a qubit for topological quantum computation. Throughout the article, we alternatively use the spin representation and Majorana fermion representation to discuss the problem. We also remind readers that although the specific physical content under these two representations differs greatly (topological edge states can only be discussed in the fermion representation, while the spin representation has no corresponding topological states), mathematically, they can be transformed into each other through Jordan-Wigner transformations.

	When the onsite dissipation ($L_{j}=-iw_{2j-1}w_{2j}$) is introduced, the aforementioned edge modes are no longer stable.
	Since the density matrix $\rho$ can be written as the combinations of $4^N$ Majorana operators $w^{\{a\}}:=w_{1}^{a_{1}}w_{2}^{a_{2}}...w_{2N}^{a_{2N}}$ with $a_j=(0,1)$, in order to find the solution of the model in this case,  we employ the third quantization formalism proposed by Prosen \cite{prosen2008third,prosen2010spectral,prosen2010exact,reslen2020uncoupled}, and vectorize the density matrix $\rho \rightarrow |\rho\rangle\rangle$ by introducing $|w^{\{a\}}\rangle \rangle$ as the basis vectors of the extended Liouville-Fock space. The master equation can then be recast  into (See Appendix A for details) a Schr\"{o}dinger-like equation
	$
	i\dot{|\rho\rangle\rangle}=\mathcal{L}|\rho\rangle\rangle,
	$
	with the corresponding non-Hermitian Liouvillian
	\bea
	\mathcal{L}&=&-2i\sum_{j=1}^{N-1}J_{j}[c_{2j}^{\dagger}c_{2j+1}+c_{2j}c_{2j+1}^{\dagger}] \nn \\
	&&   +i\sum_{j=1}^{N}\gamma_{j}[(2n_{2j-1}-1)(2n_{2j}-1)-1].  \label{eq:Liouvillian}
	\eea
	The above equation represents a dissipative spinless Hubbard model in the extended Liouville-Fock space with interlaced hoppings and interactions.
	Compared with the Hermitian case, the size of the lattice has been doubled.
	Here $c_j$ and $c^{\dag}_j$ are re-defined fermion operators in Liouville-Fock space, and satisfy the relation $\{c_i,c_j^{\dag}\}=\delta_{ij}$ and $\{c_i,c_j\}=\{c_i^{\dag},c_j^{\dag}\}=0$.
	$n_j=c_j^{\dag}c_j$ is the number operator of fermion particle on lattice $j$.
	The explicit action of $c_i$ and $c^{\dag}_i$ on the density matrix reads
	\bea
	(c_{2i-1}+c_{2i-1}^{\dagger})|\rho\rangle\rangle\rightarrow\prod_{j<i}\sigma_{j}^{z}\sigma_{i}^{x}\rho, \\
	(c_{2i}+c_{2i}^{\dagger})|\rho\rangle\rangle\rightarrow\prod_{j<i}\sigma_{j}^{z}\sigma_{i}^{y}\rho.
	\eea
	The presence of local dissipations leads to imaginary  nearest-neighboring interactions $i\gamma_j$ between the nearest lattice pairs $(2j-1,2j)$.
	Without loss of generality, we assume $J_j=J$ and $\gamma_j=\gamma (\forall j)$ in the following.
	
	The non-Hermitian  Liouvillian $\mathcal{L}$ has internal symmetry, which allows us to simplify the model significantly.
	It is easy to check for each $P_{j}=(2n_{2j}-1)(2n_{2j+1}-1)$ with $j=(1,2,\cdots,N-1)$, we have $
	[P_j, \mathcal{L}]=0 \mbox{ and } [P_j,P_k]=0$.
	Therefore the right eigenvectors of $\mathcal{L}$ can be chosen as the common eigenvectors of all $P_j$. Since $P_j^2=I$, the corresponding eigenvalues $p_j$ can only be $+1$ or $-1$. The whole Liouville-Fock space can then be divided into different subspaces labeled by the list $\{p\}=\{p_1,p_2,\cdots,p_{N-1}\}$ with $(N-1)$-entries. Since there are $2^{N-1}$ different lists, the dimension of each subspace reads $4^{N}/2^{N-1}=2 \times 2^N$. Therefore, solving the eigensystem of $\mathcal{L}$ is reduced to find all the eigenvectors of $\mathcal{L}$ within each subblock, which greatly simplifies the computation.

	\begin{figure}[ht]
		\includegraphics[scale=1]{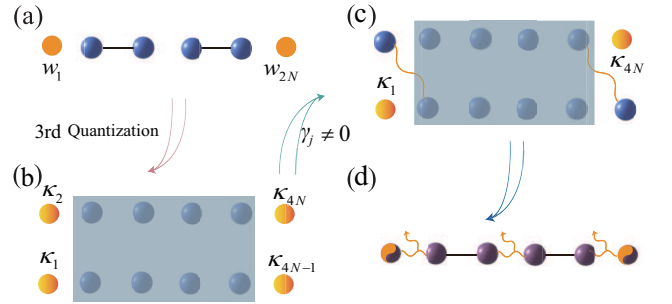}
		\caption{Diagrammatic representation of Majorana fermion (MF) and Liouville-Majorana fermion (LMF) based on the third quantization. In $(a)\rightarrow (b)$,  the Liouvullian of the system is obtained in the extended Liouville Fock space, where two isolated Hermitian MFs correspond to four isolated LMFs. Since two of these LMFs couple to the bulk modes due to dissipations with $\gamma\ne 0$, there are only two isolated LMFs $\kappa_{1},\kappa_{4N}$ in the system, as shown in $(b)\rightarrow (c)$. A LMF can be viewed as  a \textit{half-MF} after mapping back to the original Hilbert space ($(c)\rightarrow (d)$).}
		\label{fig1}
	\end{figure}

	\textit{Effective non-Hermitian spin or Kitaev chains in Liouville-Fock space.}
	To illustrate the hidden topological features of the system, we employ two cascaded Jordan-Wigner transformations  again (See Appendix A for details), and
	rewrite the  Liouvillian  $\mathcal{L}$ of the system as
	\bea
	\mathcal{L}&=&\sum_{j}^{N-1}iJ(P_{j}-1)\kappa_{4j-1}\kappa_{4j+2} \nn \\&& \hspace{1cm}+i\gamma\sum_{j}^{N}(i\kappa_{4j-2}\kappa_{4j-1}-1). \label{eq:effHmajorana}
	\eea
	Here $\{\kappa_{k}\}$ represents another new-defined set of $4N$ Liouville-Majorana fermions (LMFs) in Liouville-Fock space with $k=1,\cdots,4N$.
	The specific dependencies of $\kappa_{k}$ on $c_j$ are tedious and will not be listed here (See Appendix A and B for details).
	Therefore, within each subblock defined by $ P_{j}=i\kappa_{4j}\kappa_{4j+1} $, $\mathcal{L}$ takes the form of an effective non-Hermitian Kitaev chain with site-dependent couplings $J(p_{j}-1)$ ($2J$ or $0$) and dissipation rate $i\gamma$.
	
	Equation (\ref{eq:effHmajorana}) represents one of the main results of the current work.  Although diagonalizing $\mathcal{L}_p$ analytically for given $\{p\}$ is still difficult,  the effective coupling $J_j(p_j-1)$ vanishes when $p_j=1$. This means that the whole chain is broken at these sites. Solving the model is then reduced to the diagonalization of  $\mathcal{L}_{p}$ within each subchain, which thus greatly simplifies the calculation.
	Especially, in the subspace defined by $p_j=1$ for $1\le j \le N-1$, the effective Liouvillian is recast into $\mathcal{L}_{p} = \sum_{j=1}^{N} i\gamma_j (i\kappa_{4j-2}\kappa_{4j-1}-1) $, which describes series of isolated dissipative coupled pairs of Liouville--Majorana operators.
	The stationary states of the whole system can also be found in this subspace satisfying $i\kappa_{4j-2}\kappa_{4j-1}|\rho_s\rangle\rangle = |\rho_s\rangle\rangle $, whose general form can be written as
	$
	\rho_s = (I+ \zeta \mathcal{M})/2^N
	$
	with $\mathcal{M}=(-1)^{N}\prod_{j=1}^{N}\sigma_{j}^{z}$ and $-1 \le \zeta \le 1$.
	
	\begin{figure}[ht]
		\includegraphics[scale=1]{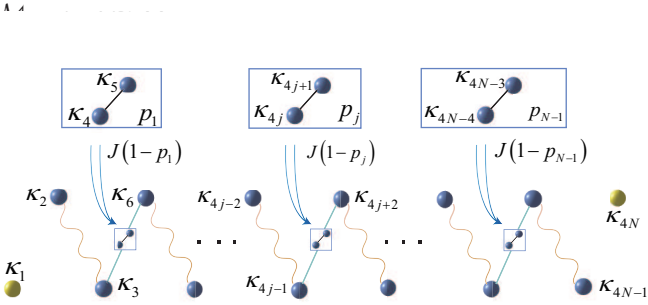}
		\caption{The reduced non-Hermitian Kitaev chain within the subspace defined by $\{p_1,p_2,\cdots,p_{N-1}\}$. For specific given $\{p\}$, this chain is broken at the lattice site satisfying $p_j=1$, and becomes an assembly of subchains with shorter length.}\label{fig:effHmajorana}
	\end{figure}

	\textit{Liouville-Majorana edge modes (LMEMs) for open boundaries.}
	For finite lattice, the system supports topological LMEMs in the extended Liouville-Fock space. Specifically, the two Liouville-Majorana modes $\kappa_1$ and $\kappa_{4N}$ are decoupled from  $\mathcal{L}$ as $
	[\mathcal{L}, \kappa_1] = [\mathcal{L}, \kappa_{4N}] = 0$.
	Therefore, we can introduce new fermion operators in this subspace $\mathcal{H}_e$ defined by $
	d_{e}=\frac{1}{2}(\kappa_{1}+i\kappa_{4N})$,   $ d_e^{\dag}=\frac{1}{2}(\kappa_{1}-i\kappa_{4N})
	$
	with the corresponding number states $|0\rangle\rangle$ and $|1\rangle\rangle$ satisfying $d^{\dag}_e|0\rangle\rangle = |1\rangle\rangle$ and $d^{\dag}_e|1\rangle\rangle=d_e|0\rangle\rangle=0$.
	This allow us to express the whole Liouville-Fock space $\mathcal{H}_{\mathcal{L}}$ as the product of two subspaces $\mathcal{H}'_{\mathcal{L}}\otimes \mathcal{H}_e$, where $\mathcal{H}'_{\mathcal{L}}$ denotes the Fock subspace expanded by other Liouville-Majorana modes $\kappa_j$ with $j=2,\cdots, (4N-1)$. Therefore, an initial product state (See Appendix C and D for the detailed constructions)
	\bea
	|\rho(0)\rangle\rangle =|\rho'\rangle\rangle \otimes (a|1\rangle\rangle + b|0\rangle\rangle) \label{eq:initialstate}
	\eea
	in the Liouville-Fock space remains unentangled during the evolution as $|\rho(t)\rangle\rangle = [\exp(-i\mathcal{L}t)|\rho'(0)\rangle\rangle] \otimes (a|1\rangle\rangle+b|0\rangle\rangle)$.

	We note that the LMEMs discussed here are very different from the conventional Majorana modes in Hermitian Kitaev chain.  Specifically, LMEMs are defined in the extended Liouville-Fock space, while the conventional Majorana edge modes are defined instead in the original Hilbert space. This ensures that LMEMs can survive in the long-time limit, while the usual Hermitian Majorana modes are unstable and decay rapidly in the presence of dissipations.  Meanwhile, in Hermitian system, the presence of topological Majorana modes enables us to define a 2-dimensional Hilbert space, where both qubit pure and mixed states can be well supported.  However, in dissipative system, although the presence of LMEMs  also enables the definition of Hilbert space in Liouville-Fock space, this does not indicate the existence of well-defined qubit subspace in the original Hilbert space defined by $H$.  Therefore, a general LMEMs can only be described as mixed states. This enables the exploration of nontrivial topological features in dissipative system based on mixed states.
	Finally, the correlation of LMEMs defined in the Liouville-Fock space does not correspond to a measurable observable directly as
	\bea
	\langle\langle\rho|i\kappa_{1}\kappa_{4N}|\rho\rangle\rangle= \tr(\rho\mathcal{M}\sigma_{1}^{x}\sigma_{N}^{x}\rho\sigma_{1}^{x}\sigma_{N}^{x}).
	\label{eq:long}
	\eea
	This correlation can always be expressed as a quadratic form of appropriately chosen observables, as will be shown in latter discussions.

	\textit{Detection of topologically protected LMEMs.}
	The presence of LMEMs can be easily manifested by considering an initial product state $|\rho(0)\rangle\rangle$ shown in Eq.(\ref{eq:initialstate}). To show this novel feature, we can choose two Hermitian operators $\{X_1,X_2\}$ such that both  $|X_1\rangle\rangle$ and $|X_2\rangle\rangle$ are product in Liouville-Fock space, and satisfy $
	|X_1\rangle\rangle = |X'\rangle\rangle |\phi_1\rangle\rangle$ and $ |X_2\rangle\rangle = |X'\rangle\rangle |\phi_2\rangle\rangle$
	with $|\phi_i\rangle\rangle$ the corresponding state vectors in $\mathcal{H}_e$.
	In the Appendixes, we have provided the explicit method of constructing all these operators $\{\rho,X_1,X_2\}$ in the original spin basis.
	Then using the identity
	\bea
	\frac{\langle X_1\rangle}{\langle X_2\rangle}=\frac{\langle\langle X_1|\rho \rangle\rangle}{\langle\langle X_2|\rho \rangle\rangle} = \frac{\langle\langle \phi_1|(a|1\rangle\rangle+b|0\rangle\rangle)}{\langle\langle \phi_2|(a|1\rangle\rangle+b|0\rangle\rangle)},
	\label{eq16}
	\eea
	we conclude that the ratio $\langle X_1 \rangle/\langle X_2 \rangle$ is time independent, and determined completely by the edge modes. However, if the edge modes and the bulk modes are coupled, or the initial state is entangled in Liouville-Fock space,  $\langle X_1 \rangle/\langle X_2 \rangle$ can be time-dependent and tends to a stable value only in the long-time limit.

\begin{figure}[ht]
		\includegraphics[scale=1]{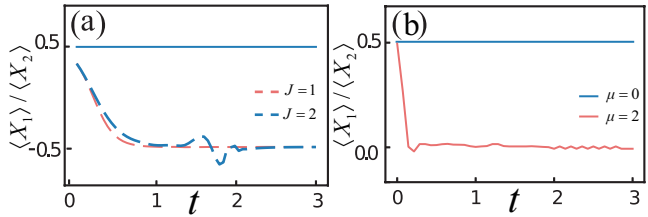}
		\caption{ (a) Evolution of the ratio  $\langle X_1 \rangle/\langle X_2 \rangle$  for initial bulk-edge product states $\rho_0$ (solid line) and non-product states $\rho_0'$ (dashed lines) in the Liouville-Fock space. The two observables are defined as $X_{1}=\sum_{j=2}^{N-1}(\sigma_{j}^{x}\sigma_{j+1}^{x}+\sigma_{j}^{z})$ and $X_2=X_1\mathcal{M}$. (b) The robustness of $\langle X_1 \rangle/\langle X_2 \rangle$ for symmetry-preserving perturbations in Eq.(\ref{eq:Hperturbation}) with randomized coupling   $\{J_j,b_j,\gamma_j,\gamma'_j\}$ for $u=0$(blue line).  $\langle X_1 \rangle/\langle X_2 \rangle$ becomes time-dependent for $u=2$(red line).  Here we have set $N=8$.
		}
		\label{fig3}
	\end{figure}

	Figure \ref{fig3} shows the evolution of the ratio defined in Eq.(\ref{eq16}) for different initial states. For initial bulk-edge product state $\rho_0=[I+\sum_{j=2}^{N-1}0.2(\sigma_{j}^{x}\sigma_{j+1}^{x}+\sigma_{j}^{z})](I+0.5\mathcal{M})/2^{N}$ and even $N$, the two observables can be chosen as  $X_{1}=\sum_{j=2}^{N-1}(\sigma_{j}^{x}\sigma_{j+1}^{x}+\sigma_{j}^{z})$ and $X_2=X_1\mathcal{M}$. The numerical calculation shows that $\langle X_1\rangle/\langle X_2\rangle$ is fixed during the evolution, as depicted by solid lines in Fig.(\ref{fig3}a). However, for non-product initial state $\rho'_0=(I+\sum_{j=2}^{N-1}0.1\sigma_{j}^{x}\sigma_{j+1}^{x})(I+0.5\mathcal{M})/2^{N}+0.1\sigma_1^z(I-0.5\mathcal{M})/2^{N}$, the dashed lines in Fig.(\ref{fig3}a) shows that the ratio $\langle X_1\rangle/\langle X_2\rangle$ changes along with $t$, which indicates the entanglement of the edge and bulk modes in this case.
	
	The edge modes are topologically protected by the internal symmetry of the system.
	For any perturbations that can be characterized by introducing additional Hamiltonian $H'$, or dissipators $L'_j$ into the Lindblad equation, the edge modes are decoupled from the bulk modes as long as the corresponding Lindbladians in Liouville-Fock space commute with  $\kappa_1$ and $\kappa_{4N}$. Using spin language, these operators can be chosen such that
	\bea
	[H'(L'),\sigma_1^x]=[H'(L'),\sigma_N^x]=[H'(L'),\prod_{j=1}^N\sigma_j^z]=0.
	\eea
	For comparative purposes, in figure (\ref{fig3}b), we also plot the evolution of $\langle X_1\rangle/\langle X_2\rangle$ for Lindblad equation with
	\bea
	H&=&\sum_{j=1}^{N-1}J_j\sigma_j^x\sigma_{j+1}^x + \sum_{j=2}^{N-1}b_j\sigma_j^z+u\sum_{j=1}^N\sigma_j^x, \label{eq:Hperturbation}
	\eea
	the dissipator $L_j=\gamma_j\sigma_j^z (j=1,\cdots,N)$ and  $L'_j=\gamma'_j \sigma_j^x\sigma_{j+1}^x  (j=1,\cdots,N-1)$. For the initial bulk-edge product state $\rho_{0}$, the calculation shows that $\langle X_1\rangle/\langle X_2\rangle$ remains fixed for all coefficients $\{J_j,b_j,\gamma_j,\gamma'_j\}$ randomly distributed between $0$ and $1$ when $u=0$, which proves the robustness of edge modes under symmetry-preserving perturbations. For nonzero $u=2$, $\langle X_1\rangle/\langle X_2\rangle$ changes along with $t$ as the edge modes couples to the bulk due to the perturbations.

	\textit{ Purity as the detection of Long-range correlation in Liouville-Fock space. }
	For the initial state $|\rho(0)\rangle\rangle$ shown in Eq.(\ref{eq:initialstate}), since $i\kappa_1\kappa_{4N} |\rho\rangle\rangle = |\rho'\rangle\rangle \otimes (a|1\rangle\rangle-b|0\rangle\rangle)$, the correlation defined by  $\langle\langle \rho|i\kappa_1\kappa_{4N}|\rho \rangle\rangle$ can be written as
	\bea
	\langle\langle \rho | i\kappa_1\kappa_{4N} |\rho\rangle\rangle = \frac{|a|^2-|b|^2}{|a|^2+|b|^2} \langle \langle \rho|\rho \rangle\rangle \propto \Tr(\rho^2),
	\eea
	where $\Tr(\rho^2)=\langle \langle \rho|\rho \rangle\rangle$ is the purity of the state $\rho$. After inserting the completeness relation in Liouville-Fock space, we have
	\bea
	\langle \langle \rho|\rho \rangle\rangle=\sum_{\mu}\frac{\langle \langle \rho|O_{\mu}\rangle\rangle\langle\langle O_{\mu}| \rho\rangle\rangle}{2^N}  =\sum_{\mu} \frac{|\langle O_{\mu}\rangle|^2}{2^N}   \label{eq:observ}
	\eea
	with $O_{\mu}$ the usual $N$-spin Pauli operators (See In the Appendix E for details).
	Hence, the correlation $\langle\langle \rho|i\kappa_1\kappa_{4N}|\rho \rangle\rangle$ can be expressed as a quadratic form of observables defined by $O_{\mu}$.
	For dissipative systems, the dynamics in the long time limit is mainly determined by eigenvectors in the expansion of $|\rho(t)\rangle\rangle=\sum_{j} e^{-i\lambda_{j} t}|\rho'_{j}\rangle\rangle \otimes (a|1\rangle\rangle+b|0\rangle\rangle)$ with minimal $|\Im(\lambda_{j})|$ as $\Im(\lambda_{j})\le0$. This indicates that
	the summation in Eq.(\ref{eq:observ}) can be well approximated by choosing a subset $\{O_m (m=1,\cdots,M)\}$ with many fewer  observables  ($M\ll N$), which can then simplify detection in experiment.

	\begin{figure}[ht]
		\includegraphics[scale=1]{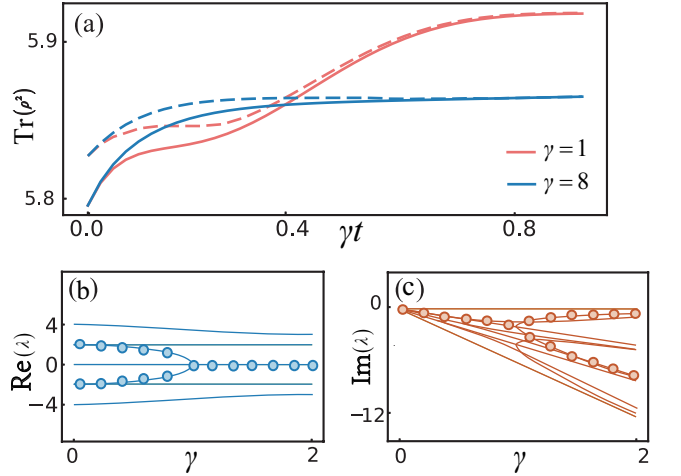}
		\caption{(a) Evolution of the purity $\Tr(\rho^2)$ for an initial bulk-edge product state. The dotted line is the approximated purity obtained by Eq.(\ref{eq:renyi}), which is  compared with the precise values obtained by solving the Lindblad equation directly (solid line). The two results match well for larger $\gamma t$.
			(b) and (c) are the real and imaginary parts of Liouville spectrum for $(N=6,J=2)$. The circles are the eigenvalues within the subspace defined by $\{p_{j\ne2}=1(\forall j),p_{2}=-1\}$.  }
		\label{fig4}
	\end{figure}

	In Fig. \ref{fig4}, we plot the evolution of $\langle \langle \rho|\rho \rangle\rangle$ for the initial product state $\rho_0=[I+0.3\mathcal{M}(\sigma_1^z+\sigma_1^y\sigma_2^x+\sigma_1^y\sigma_3^x+\sigma_2^z\sigma_2^x\sigma_3^x)](I+0.4\mathcal{M})/2^{N}$.
	This state has nonzero components in subspaces defined by $\{p_j=1 (\forall j)\}$ and $\{p_{j\ne1}=1(\forall j),p_{1}=-1\}$.
	The Lindblad spectra $\{\lambda_i\}$ show exceptional points as we increase the dissipation rate $\gamma$, as shown in Fig.  \ref{fig4}b and \ref{fig4}c.
	In addition, the system supports numerous quasi-stable state $\Im(\lambda) \rightarrow 0^-$ as $\gamma \rightarrow \infty$.
	The correlation  $\langle\langle \rho | i\kappa_1\kappa_{4N} |\rho\rangle\rangle$ can then be obtained as
	\bea
	\langle \langle \rho|i\kappa_{1}\kappa_{4N}|\rho\rangle\rangle \sim \Big(1+\langle\sigma_{1}^{y}\sigma_{2}^{x}\rangle^{2}+\langle\sigma_{1}^{z}\rangle^{2}\Big).   \label{eq:renyi}
	\eea
	Hence the long-time evolution of purity for the given state can be obtained by detecting only short-range correlations defined by $\langle\sigma_{1}^{y}\sigma_{2}^{x}\rangle$ and $\langle\sigma_{1}^{z}\rangle$. For larger decay rate $\gamma$, the result fits the exact result well, as shown by dashed line in Fig.  \ref{fig4}a.

	\textit{Discussion and conclusion.}To summarize, by solving an exactly solvable model of open system described by Lindblad master equations, we find a topologically protected Liouville-Majorana modes hidden in the Liouvillian.
	We proved that generally, the mode corresponds to mixed states of the system, which is different from the case in Hermitian system, where it can be described in terms of pure states.
	The mode is robust and stable in the whole dynamic process, which is also different from the stationary state of the Liouville equation.
	The work opens up the research of nontrivial topological states defined in the extended Liouville-Fock space and extends the exploration of topological physics for mixed states in general dissipative systems.
	

	We thank Prof. X.-W. Luo for helpful discussions. This work was funded by National Natural Science Foundation of China (Grants No. 11974334, and No. 11774332), and Innovation Program for Quantum Science
	and Technology (Grant No. 2021ZD0301200).XFZ also acknowledges support from CAS Project for Young Scientists in Basic Research (Grant No.YSBR-049).

	
	\vspace{0.2cm}
		
	\begin{widetext}
	\appendix
	\vspace{0.2cm}
	In this Appendix, we present the explicit derivation of the effective Liouvillian in the extended Liouville Fock space using the third quantization formalism. Furthermore, the explicit construction of the bulk-edge product vectors in Liouville-Fock space is provided, and the relevant forms in the original spin basis are discussed. Finally, the construction methods and concrete forms of probe operators are discussed in detail.
	
	\section{ \label{A} Liouville-Fock space and Non-Hermitian effective Liouvillian based on Prosen's third quantization}
	
	For open system with Markov approximation, the dynamics of its density
	matrix $\rho$ is governed by the following Lindblad master equation
	\bea
	i\dot{\rho}=\hat{\phi}_{L}[\rho]=[H,\rho]+i\sum_{j}(L_{j}\rho L_{j}^{\dagger}-\frac{1}{2}\{L_{j}^{\dagger}L_{j},\rho\})
	\eea
	which describes the non-unitary time evolution of the system subject
	to the external environment. Here the first term represents the unitary
	dynamics where $H$ is the Hamiltonian of the system. $L_{j}$ is
	the corresponding Lindblad operator describing the $j$-th dissipation
	channel with the decay rate $\gamma_{j}$.
	
	By regarding $\rho$ as a vector $|\rho\rangle\rangle$ and due to the linearity
	of the system, we can rewrite the equation as
	\bea
	i\dot{|\rho\rangle\rangle}=\mathcal{L}|\rho\rangle\rangle,
	\eea
	which takes similar form as the usual Schrondinger equation with
	the effective non-Hermitian Liouvillian $\mathcal{L}$. Generally
	speaking, the explicit form of $\mathcal{L}$ depends on how we vectorize
	the matrix $\rho$. Specifically, for quadratic spin/fermi system,
	the vectorization  process can be easily discussed based on Majorana
	representation. Especially, Prosen has introduced the third quantization
	formalism in \cite{prosen2008third,prosen2010spectral,prosen2010exact}, which allow us to solve this dissipated system
	in an elegant and systematic manner.
	
	For $N$-spin/fermi system  in 1D, the corresponding density matrix can be
	written using Majorana operators as
	\bea
	\rho=\frac{1}{2^{2N}}\sum c_{a_{1},a_{2},a_{3}...a_{2N}}w_{1}^{a_{1}}w_{2}^{a_{2}}w_{3}^{a_{3}}...w_{2N}^{a_{2N}}
	\eea
	where $w_{j}$ are Majorana operators satisfying the anti-commutation
	relation $\{w_{j},w_{k}\}=2\delta_{jk}$,$a_{j}=\{0,1\}$ represents
	the excitation number of the $w_{j}$, and $c_{a_{1},a_{2},a_{3}...a_{2N}}$
	are the real coefficients. For spin-$1/2$ system discussed in the main text, this
	is always possible due to the Jordan-Wigner transformation
	\bea
	\sigma_{j}^{x}&=&\prod_{k<j}(-iw_{2k-1}w_{2k})w_{2j-1}, \\
	\sigma_{j}^{y}&=&\prod_{k<j}(-iw_{2k-1}w_{2k})w_{2j}.
	\eea
	
	For later convenience, we define
	\bea
	w^{\{a\}}:=w_{1}^{a_{1}}w_{2}^{a_{2}}...w_{2N}^{a_{2N}},
	\eea
	and $n_{a}:=\sum_{j}a_{j}$ represents the number of Majorana operators
	in the basis vector $|w^{\{a\}}\rangle\rangle$. The vectorization of the
	master equation can be implemented by associating a Hilbert space
	$\mathcal{H}_{\mathcal{L}}$, namely Liouville-Fock space, with the basis defined as
	\bea
	|w^{\{a\}}\rangle\rangle:=|w_{1}^{a_{1}}w_{2}^{a_{2}}...w_{2N}^{a_{2N}}\rangle\rangle.
	\eea
	to the $4^{n}$-dimensional space of operators $w^{\{a\}}$. In the
	meantime, since the Hamiltonian and relevant Lindbladians can also
	be written as the combinations of Majorana operators $w^{\{a\}}$,
	the Liouville superoperator $\mathcal{L}$ can then be expressed as
	an operator in this newly-defined Liouville-Fock space. Specifically, for
	each Majorana operator $w_{k}$ shown in $H$ or $L_{i}$ acting on
	the basis $|w^{\{a\}}\rangle\rangle$, we can introduce the fermions operators
	$c_{j}$ and $c_{j}^{\dagger}$ as
	\bea
	c_{j}^{\dagger}|w_{1}^{a_{1}}w_{2}^{a_{2}}...w_{2N}^{a_{2N}}\rangle\rangle=\delta_{0,a_{j}}|w_{j}w_{1}^{a_{1}}w_{2}^{a_{2}}...w_{2N}^{a_{2N}}\rangle\rangle,
	\eea
	
	\bea
	c_{j}|w_{1}^{a_{1}}w_{2}^{a_{2}}...w_{2N}^{a_{2N}}\rangle\rangle=\delta_{1,a_{j}}|w_{j}w_{1}^{a_{1}}w_{2}^{a_{2}}...w_{2N}^{a_{2N}}\rangle\rangle,
	\eea
	with the standard canonical anticommutation relations
	\bea
	\{c_{j},c_{k}\}=0,\{c_{j},c_{k}^{\dagger}\}=\delta_{jk},\{c_{j}^{\dagger},c_{k}^{\dagger}\}=0.
	\eea
	For $N$-site spin/fermi system, since the dimension of the Fock
	space $\mathcal{H}_{\mathcal{L}}$ is $4^{N}$, we have $2N$ fermion operators $c_j$ with $j=(1,2,\cdots,2N)$.
	
	The master equation of the system can be written using majorana operators as
	\bea
	i\dot{|\rho\rangle\rangle}=-i\sum_{j=1}^{N-1}J_j(w_{2j}w_{2j+1}\rho-\rho w_{2j}w_{2j+1})+i\sum_{j=1}^{N }\gamma_j(w_{2j-1}w_{2j}\rho w_{2j}w_{2j-1}-\rho).
	\eea
	Based on the above discussions, one can verify that after mapping into the Liouville-Fock space, operators acting on $\rho$ can be re-expressed using fermionic operators as
	\bea
	\omega_j\rho  &\Longrightarrow& (c_j+c_j^{\dag})|\rho\rangle\rangle, \\ \rho\omega_i\omega_j &\Longrightarrow& (c_j-c_j^{\dag})(c_j-c_j^{\dag})|\rho\rangle\rangle.
	\eea
	Using these substitutions, we can immediately obtain the relevant Liouvillian $\mathcal{L}$ which reads
	\bea
	\mathcal{L}&=&-2i\sum_{j=1}^{N-1}J_j(c_{2j}^{\dagger}c_{2j+1}+c_{2j}c_{2j+1}^{\dagger}) -i\sum_{j=1}^N\gamma_j  +i\sum_{j=1}^{N}\gamma_j(2n_{2j-1}-1)(2n_{2j}-1),
	\eea
	where $n_j=c^{\dag}_jc_j$ is the number operator on site $j$.
	Since  $\mathcal{L}$ commutes with all $P_{j}=(2n_{2j}-1)(2n_{2j+1}-1)$ for $j=(1,2,\cdots,N-1)$, and $P_j^2=I$, the right eigenvectors of $\mathcal{L}$ can be chosen as the common eigenvectors of all $P_j$, where the corresponding eigenvalues $p_j$ can only be $+1$ or $-1$. The whole Liouville-Fock space can then be divided into different subspaces labeled by the list $\{p\}=\{p_1,p_2,\cdots,p_{N-1}\}$ with $(N-1)$-entries.
	
	To obtain the effective interactions of  $\mathcal{L}$, we introduce another set of Jordan-Wigner transformations (JW-I) in Liouville-Fock space as $ c_{2i-1}^{\dagger}=\frac{1}{2}\prod_{j=1}^{2i-2}Z_{j}(X_{2i-1}-iY_{2i-1})$ and $c_{2i}^{\dagger}=\frac{1}{2}\prod_{j=1}^{2i-1}Z_{j}(Y_{2i}-iX_{2i})$, and map the system into an effective spin model defined as
	\bea
	\mathcal{L}=\sum_{j}^{N-1}J(P_{j}-1)Y_{2j}Y_{2j+1}-i\gamma\sum_{i}^{N}(Z_{2j-1}Z_{2j}+1).  \label{eq:effHspin}
	\eea
	Here $\{X_k,Y_k,Z_k\}$ are the local Pauli matrices defined in Liouville-Fock space at site $k$, and we have set the homogeneous decay rates as $\gamma_j=\gamma$. Therefore, within each subblock denoted by $\{p\}$, $\mathcal{L}$ takes the form of a non-Hermitian spin mode with site-dependent couplings $J(p_{j}-1)$ and dissipation rate $i\gamma$.
	
	To illustrate the hidden topological features of the system, we employ  the Jordan-Wigner transformation (JW-II)  again and define the local Liouville-Majorana operators as $\kappa_{2i-1}=-\prod_{j=1}^{i-1}X_{j}Z_{i}$ and $ \kappa_{2i}=\prod_{j=1}^{i-1}X_{j}Y_{i}$, and finally we arrive at
	\bea
	\mathcal{L}=\sum_{j}^{N-1}iJ(P_{j}-1)\kappa_{4j-1}\kappa_{4j+2}+i\gamma\sum_{j}^{N}(i\kappa_{4j-2}\kappa_{4j-1}-1) 
	\eea
	with $P_{j}=i\kappa_{4j}\kappa_{4j+1}$.
	Therefore, for given $\{p\}$, $\mathcal{L}$ redueces to an effective non-Hermitian Kitaev chain with site-dependent couplings.
	
	We stress that although both $\omega_j$ and $\kappa_j$ are Majorana operators (MOs), they are defined in different spaces.
	Specifically, $\omega_j$ is the MO defined in the original Hilbert space, and $\kappa_j$ is another type of MO defined in the extended Liouville-Fock space (denoted by $\mathcal{H}_{\mathcal{L}}$ in the paper).
	For $N$-site chain, we have $2N$ $\omega$-type MOs, but $4N$ $\kappa$-type Liouville-MOs. So generally speaking, one $\omega$-type MO maps to two $\kappa$-type MOs.
	In this sense, we claim that a Liouville-Majorana fermion can be viewed as a half-Majorana fermion in the original Hilbert space. In the spin basis defined in Eq.(1), the Liouville-Majorana edge modes discussed in the paper can only be described as mixed states, which is different from the case for the usual Majorana mode.

	\section{ \label{B} The spectra and dynamical features of Liouvillian $\mathcal{L}$}
	
	To explore the dynamical properties of system, we consider the eigenmatrices
	and eigenvalues of the Liouville superoperator $\hat{\phi}_{L}$ and
	its counterpart $\mathcal{L}$ in the Liouville  Fock space $\mathcal{H}_{\mathcal{L}}$
	\bea
	\hat{\phi}_{L}[\rho_{m}]=\lambda_{m}\rho_{m}\rightarrow\mathcal{L}|\rho_{m}\rangle\rangle=\lambda_{m}|\rho_{m}\rangle\rangle.
	\eea
	For a master equation in the Lindblad form, it has been shown that
	the spectrum $\{\lambda_{m}\}$ satisfies the following properties
	which are useful for later discussions.
	
	First, since the imaginary
	part of $\lambda_{m}$ is linked with the dissipation dynamics towards
	stationary states, we have $\Im[\lambda_{m}]\le0$. The stationary state $\rho_{ss}$
	of the system corresponds to the eigenmatrix $\rho_{0}$ with $\lambda_{0}=0$.
	So we have $\rho_{ss}=\rho_{0}/\mbox{Tr}[\rho_{0}]$. Additionally,
	if stationary states are degenerate, the system can evolve towards different
	steady states depending on the initial conditions.
	
	Second, since $\rho$
	is Hermitian, and $\hat{\phi}_{L}[\sigma^{\dagger}]=-(\hat{\phi}_{L}[\sigma])^{\dagger}$
	for any matrix $\sigma$, the eigenvalues must come in anti-complex
	conjugate pairs $\{\lambda_{m},-\lambda_{m}^{*}\}$. Therefore, if
	$\lambda_{m}$ is pure imaginary, the eigenmatrix $\rho_{m}$ must
	be Hermitian and vice versa.
	
	Finally, if $\Im[\lambda_{m}]\neq0$,
	since the Liouvillian evolution is trace-preserving, the eigenmatrix
	evolves as $e^{-i\lambda_{m}t}\rho_{m}\rightarrow0$ when $t\rightarrow\infty$.
	This leads to $\mbox{Tr}[\rho_{m}]=0$.
	
	Equipped with the eigensystem of the Lindblad equation, we can then
	discuss the dynamics of the system in a more convenient manner. Since
	any physical state of the system can always be decomposed as
	\bea
	\rho=g_{0}\rho_{ss}+\sum_{m\neq0}g_{m}\rho_{m},
	\eea
	the time-evolution of $\rho(t)$ in Liouville-Fock space $\mathcal{H}_{\mathcal{L}}$ can then be simplified as
	\bea
	|\rho(t)\rangle\rangle=g_{0}|\rho_{ss}\rangle\rangle+\sum_{m\neq0}g_{m}e^{-i\lambda_{m}t}|\rho_{m}(t)\rangle\rangle.
	\label{eq:time evl}
	\eea
	
	We stress that the dynamical properties of a quantum system with the effective Liouvillian $\mathcal{L}$ is very different from the usual
	non-Hermitian system solely driven by an effective non-Hermitian Hamiltonian
	$H_{e}=H-i\gamma\sum_{m}L_{m}^{\dagger}L_{m}/2$. In the later case,
	the effect of quantum jump $L_{m}\rho L_{m}^{\dagger}$ has been neglected.
	We also note that the non-Hermiticity of $H_{e}$ can result in many novel effects.
	For instance, pseudo-Hermitian or PT-symmetric $H_{e}$
	has been widely discussed in the past decades, which gives rise to
	rich exotic phenomena in different subjects of physics. However, in
	many cases, this jump term $L_{m}\rho L_{m}^{\dagger}$ cannot be
	dropped and can change the dynamical behavior of the system dramatically.

	\section{\label{C} Bulk-edge product vectors in Liouville-Fock space}
	
	In our system, the two edge Liouville-Majorana operators $\kappa_{1}$ and $\kappa_{4N}$ can be used to define the Dirac fermionic operator $d_{e}=\frac{1}{2}(\kappa_{1}+i\kappa_{4N})$ and $d^{\dag}_{e}=\frac{1}{2}(\kappa_{1}-i\kappa_{4N})$.
	The corresponding number operator reads $d_{e}^{\dagger}d_{e}$ and satisfies the following properties after acting on its local Fock basis
	\bea
	d_{e}^{\dagger}d_{e}|1\rangle\rangle=|1\rangle\rangle, \hspace{1.cm} d_{e}^{\dagger}d_{e}|0\rangle\rangle=0.
	\eea
	Since $d_{e}$ and $d^{\dag}_{e}$ commute with the Liouvillian $\mathcal{L}$, $|0\rangle\rangle$ and $|1\rangle\rangle$ correspond to the two local dark states of the system, and defined as the basis of the local Fock space denoted by $\mathcal{H}_{e}$.
	For the remaining  Liouville-Majorana operators $\kappa_{j}$ with $2\le j\le (4N-1)$, they can  be combined similarly to define $2N-1$ Dirac fermionic operators with the corresponding  Fock space denoted by $\mathcal{H}'_{\mathcal{L}}$.
	Therefore, the whole Liouville-Fock space $\mathcal{H}_{\mathcal{L}}$ can then be expressed as the tensor product of $\mathcal{H}'_{\mathcal{L}}$ and $\mathcal{H}_{e}$.
	Using these notations, we can then rewrite the state vector $|\rho\rangle\rangle$ in Liouville-Fock space as
	\bea
	|\rho\rangle\rangle=|\psi_{1}\rangle\rangle|1\rangle\rangle+|\psi_{0}\rangle\rangle|0\rangle\rangle.
	\eea
	
	For operators acting on $|\rho\rangle\rangle$ in Liouville-Fock space, they can be mapped to the corresponding linear operations in the original Hilbert space defined by the spin basis.
	For later convenience, we list the explicit correspondence as follows
	\bea
	d_{e}^{\dagger}d_{e}|\rho\rangle\rangle &\rightarrow &\frac{1}{2}(\rho+\mathcal{M}\sigma_{1}^{x}\sigma_{N}^{x}\rho\sigma_{1}^{x}\sigma_{N}^{x}),\\
	d_{e}|\rho\rangle\rangle &\rightarrow &\frac{1}{2}\mathcal{M}\sigma_{1}^{x}(\mathcal{M}\sigma_{1}^{x}\sigma_{N}^{x}\rho\sigma_{N}^{x}\sigma_{1}^{x}+\rho)\mathcal{M}\sigma_{1}^{x},\\
	d_{e}^{\dagger}|\rho\rangle\rangle &\rightarrow &-\frac{1}{2}\mathcal{M}\sigma_{1}^{x}(\mathcal{M}\sigma_{1}^{x}\sigma_{N}^{x}\rho\sigma_{N}^{x}\sigma_{1}^{x}-\rho)\mathcal{M}\sigma_{1}^{x}.
	\eea
	where $\mathcal{M}=(-1)^{N}\prod_{j=1}^{N}\sigma_{j}^{z}$.
	One can check that if $|\rho\rangle\rangle=|\rho_1\rangle\rangle=|\psi_{1}\rangle\rangle|1\rangle\rangle$, then we have
	\bea
	d_{e}^{\dagger}d_{e}|\rho_1\rangle\rangle=|\rho_1\rangle\rangle\rightarrow\mathcal{M}\sigma_{1}^{x}\sigma_{N}^{x}\rho_1\sigma_{1}^{x}\sigma_{N}^{x}=\rho_1.
	\label{d6}
	\eea
	Similarly, if $|\rho\rangle\rangle=|\rho_{0}\rangle\rangle=|\psi_{0}\rangle\rangle|0\rangle\rangle$, we have
	\bea
	d_{e}^{\dagger}d_{e}|\rho_{0}\rangle\rangle=|\rho_{0}\rangle\rangle\rightarrow\mathcal{M}\sigma_{1}^{x}\sigma_{N}^{x}\rho_0\sigma_{1}^{x}\sigma_{N}^{x}=-\rho_0.
	\label{d7}
	\eea
	This also indicates that if $\rho_i$ is hermitian, then we must have $[\rho_i,\prod_{j}^{N}\sigma_{j}^{z}]=0$.
	
	To obtain the explicit form in Liouville-Fock space for a given density matrix, we consider the following $N$-body Pauli operator in the original Hilbert space  $\hat{O}=\sigma_{1}^{\mu_{1}}\otimes\sigma_{2}^{\mu_{2}}\otimes...\otimes\sigma_{N}^{\mu_{N}}$
	with $\mu_{i}\in\{0,x,y,z\}$ and $\sigma^0= I$ the usual identity matrix.
	The relevant state vector in Liouville-Fock space reads
	\bea
	|\hat{O}\rangle\rangle=|\hat{O}_{1}\rangle\rangle|1\rangle\rangle+|\hat{O}_{0}\rangle\rangle|0\rangle\rangle.
	\label{d8}
	\eea
	In order to show that $|\hat{O}\rangle\rangle$ can be written as a product state in Liouville-Fock space, we define the following two projectors
	\[
	P_{+}=\frac{1}{2}(I+\mathcal{M}),\,\, P_{-}=\frac{1}{2}(I-\mathcal{M})
	\]
	with $P^2_{\pm}=I$.
	Since $\hat{O}$ is commuted ($\delta_o=+1$) or anti-commuted ($\delta_o=-1$) with $\prod_{i}^{N}\sigma_{i}^{z}\sigma_{1}^{x}\sigma_{N}^{x}$ as
	\bea
	\hat{O}\prod_{i=1}^{N}\sigma_{i}^{z}\sigma_{1}^{x}\sigma_{N}^{x}=\delta_{o}\prod_{i=1}^{N}\sigma_{i}^{z}\sigma_{1}^{x}\sigma_{N}^{x}\hat{O},
	\eea
	we have
	\bea
	\mathcal{M}\sigma_{1}^{x}\sigma_{N}^{x}\hat{O}P_{\pm}\sigma_{1}^{x}\sigma_{N}^{x}=\delta_{o}\hat{O}P_{\pm}\mathcal{M}=\delta_{o}\hat{O}P_{\pm}. \nn\\
	\eea
	Using Eq.(\ref{d6}) and (\ref{d7}), we conclude that $|\hat{O}P_{+}\rangle\rangle$ and $|\hat{O}P_{-}\rangle\rangle$ can be written as
	\bea
	|\hat{O}P_{\pm}\rangle\rangle=|\hat{O}_{\pm}\rangle\rangle|\frac{1\pm\delta_{o}}{2}\rangle\rangle,
	\eea
	where $|\hat{O}_{\pm}\rangle\rangle$ represent the corresponding state vectors in $\mathcal{H}'_{\mathcal{L}}$, whose explicit forms are irrelevant to the latter discussion.
	Therefore, the vectors related to $\hat{O}$ and $\hat{O}\prod_{i}^{N}\sigma_{i}^{z}$ then read
	\bea
	|\hat{O}\rangle\rangle &=&|\hat{O}_{+}\rangle\rangle|\frac{1+\delta_{o}}{2}\rangle\rangle+|\hat{O}_{-}\rangle\rangle|\frac{1-\delta_{o}}{2}\rangle\rangle, \\
	|\hat{O}\mathcal{M}\rangle\rangle&=&|\hat{O}_{+}\rangle\rangle|\frac{1+\delta_{o}}{2}\rangle\rangle-|\hat{O}_{-}\rangle\rangle|\frac{1-\delta_{o}}{2}\rangle\rangle.
	\eea
	
	In order to show that both $|\hat{O}\rangle\rangle$ and $|\hat{O}\mathcal{M}\rangle\rangle$ can be written as product vectors in Liouville-Fock space, we need to show that $|\hat{O}_{+}\rangle\rangle\propto|\hat{O}_{-}\rangle\rangle$.
	This can be achieved by noticing that
	\bea
	d_{e}|\hat{O}P_{+}\rangle\rangle=\frac{1+\delta_{o}}{2}|\hat{O}_{+}\rangle\rangle|0\rangle\rangle,
	\eea
	which is non-zero only when $\delta_o=+1$.
	The corresponding matrix form in the original Hilbert space reads
	\bea
	&&-\frac{1}{2}\mathcal{M}\sigma_{1}^{x}(\mathcal{M}\sigma_{1}^{x}\sigma_{N}^{x}\hat{O}P_{+}\sigma_{N}^{x}\sigma_{1}^{x}+\hat{O}P_{+})\sigma_{1}^{x}\mathcal{M}  =-\frac{1+\delta_{o}}{2}\sigma_{N}^{x}\hat{O}P_{+}\sigma_{N}^{x}=-\frac{1+\delta_{o}}{2}\sigma_{N}^{x}\hat{O}\sigma_{N}^{x}P_{-}.
	\eea
	By setting $\delta_o=+1$ and noticing $\sigma_{N}^{x}\hat{O}=\gamma_o\sigma_{N}^{x}\hat{O}$ with $\gamma_o=\pm1$, we have
	\bea
	|\hat{O}_{+}\rangle\rangle|0\rangle\rangle=-\gamma_{o}|\hat{O}_{-}\rangle\rangle|0\rangle\rangle,
	\eea
	which leads to $|\hat{O}_{+}\rangle\rangle=-\delta_o\gamma_o|\hat{O}_{-}\rangle\rangle$.
	
	We note that similar result can also be obtained if we consider
	\bea
	d_{e}^{\dagger}|\hat{O}P_{+}\rangle\rangle=\frac{1-\delta_{o}}{2}|\hat{O}_{+}\rangle\rangle|1\rangle\rangle,
	\eea
	for $\delta_o=-1$.
	The corresponding matrix form reads
	\bea
	&&-\frac{1}{2}\mathcal{M}\sigma_{1}^{x}(-\mathcal{M}\sigma_{1}^{x}\sigma_{N}^{x}(\hat{O}+\hat{O}\mathcal{M})\sigma_{N}^{x}\sigma_{1}^{x}+\hat{O}+\hat{O}\mathcal{M})\sigma_{1}^{x}\mathcal{M}=-\frac{\delta_{o}-1}{2}\sigma_{N}^{x}\hat{O}P_{+}\sigma_{N}^{x}=-\frac{\delta_{o}-1}{2}\gamma_{o}\hat{O}P_{-}.
	\eea
	After writing back to the Liouville-Fock space, we again obtain $|\hat{O}_{+}\rangle\rangle=-\delta_{o}\gamma_{o}|\hat{O}_{-}\rangle\rangle$.
	
	Summing up all the above discussions, we conclude that both  $|\hat{O}\rangle\rangle$ and $|\hat{O}\mathcal{M}\rangle\rangle$ are product and read
	\bea
	|\hat{O}\rangle\rangle&=&|\hat{O}_{+}\rangle\rangle\bigg[|\frac{1+\delta_{o}}{2}\rangle\rangle-\delta_{o}\gamma_{o}|\frac{1-\delta_{o}}{2}\rangle\rangle\bigg],  \\
	|\hat{O}\mathcal{M}\rangle\rangle&=&|\hat{O}_{+}\rangle\rangle\bigg[|\frac{1+\delta_{o}}{2}\rangle\rangle+\delta_{o}\gamma_{o}|\frac{1-\delta_{o}}{2}\rangle\rangle\bigg].
	\eea
	
	where other relevant coefficients are defined as follows
	\bea
	\sigma_{N}^{x}\hat{O}&=&\gamma_o\sigma_{N}^{x}\hat{O},  \label{eq:matrixcond1}
	\\
	\hat{O}\prod_{i}^{N}\sigma_{i}^{z}\sigma_{1}^{x}\sigma_{N}^{x}&=&\delta_o\prod_{i}^{N}\sigma_{i}^{z}\sigma_{1}^{x}\sigma_{N}^{x}\hat{O}.  \label{eq:matrixcond2}
	\eea
	The above derivation also indicates that the $4^N$ operators $\hat{O}_{\mu}$ can be divided into $2^{2N-1}$ different pairs up to a constant phase factors as $(\hat{O}_{\mu},\hat{O}_{\mu}\mathcal{M})$. For any two different pairs $(\hat{O}_1,\hat{O}_1\mathcal{M})$ and $(\hat{O}_2,\hat{O}_2\mathcal{M})$, since
	\bea
	\tr(\hat{O}_i\hat{O}_j\mathcal{M})=0, \hspace{0.25cm}  \tr(\mathcal{M}\hat{O}_i\hat{O}_j\mathcal{M})=2^N\delta_{ij},
	\eea
	we have
	\bea
	\langle \langle \hat{O}_i | \hat{O}_j \rangle\rangle = 2^N \delta_{ij}, \hspace{0.25cm} \langle \langle \hat{O}_{i,+} | \hat{O}_{j,+} \rangle\rangle = 2^{N-1} \delta_{ij}.
	\eea

	Given the state vector in Liouville-Fock space shown as Eq.(\ref{d8}), we also can easily obtain the matrix form in the usual Hilbert space using the following maps
	\bea
	\delta_o&=&+1 :  \left \{ \begin{array}{l} |\hat{O}_{+}\rangle\rangle|1\rangle\rangle\rightarrow\hat{O}P_{+},\\
		|\hat{O}_{+}\rangle\rangle|0\rangle\rangle\rightarrow-\gamma_{o}|\hat{O}_{-}\rangle\rangle|0\rangle\rangle=-\gamma_{o}\hat{O}P_{-},
	\end{array}  \right. \\
	\delta_o&=&-1 :  \left \{ \begin{array}{l} |\hat{O}_{+}\rangle\rangle|0\rangle\rangle\rightarrow\hat{O}P_{+}, \\
		|\hat{O}_{+}\rangle\rangle|1\rangle\rangle\rightarrow\gamma_{o}|\hat{O}_{-}\rangle\rangle|0\rangle\rangle=\gamma_{o}\hat{O}P_{-}.
	\end{array}  \right.
	\eea

	\section{ \label{D} Bulk-edge product states in Liouville-Fock space and the corresponding density operators in the original Hilbert space}
	
	For the system consider in the main text, the general form of the stationary states $\rho_s$ can be written as the combination of $(\hat{O},\hat{O}\mathcal{M})$ with $\hat{O}=I$ and $\delta_o=\gamma_o=1$. This means
	\bea
	\rho_s = \frac{1}{2^{N}}(I+\zeta\mathcal{M})=\frac{1}{2^{N}}\bigg[(1+\zeta)IP_{+}+(1-\zeta)IP_{-}\bigg], 
	\eea
	where $\zeta$ is real and satisfies $|\zeta|\le 1$ to ensure the positivity of $\rho_s$. The corresponding vector in Liouville-Fock space reads
	\bea
	|\rho_s\rangle\rangle = \frac{1}{2^{N}}|I_{+}\rangle\rangle \bigg[(1+\zeta)|1\rangle\rangle-(1-\zeta)|0\rangle\rangle \bigg]. \label{eq:stationarystate}
	\eea
	
	For a given initial state vector $|\rho(t=0)\rangle\rangle$ in Liouville-Fock space, if $|\rho(t=0)\rangle\rangle = |\rho'\rangle\rangle \otimes (a|1\rangle\rangle + b|0\rangle\rangle)$ is product, then the state vector $|\rho(t)\rangle\rangle$ remains unentangled in Liouville-Fock space under time evolution.
	Since the system tends to its stationary state defined by Eq.(\ref{eq:stationarystate}), we conclude that the product state can always be rewritten as
	\bea
	|\rho\rangle\rangle=|\rho_+\rangle\rangle\otimes[(1+\zeta)|1\rangle\rangle-(1-\zeta)|0\rangle\rangle]/2^{N},
	\eea
	where $|\rho_+\rangle\rangle$ can be written as
	\bea
	|\rho_+\rangle\rangle=|I_+\rangle\rangle+\sum\chi_{m}|\hat{O}_{m,+}\rangle\rangle,
	\eea
	where both the coefficients $\chi_{m}$ and operators $\hat{O}_{m}$ should be carefully chosen so that corresponding $\rho$ in the original Hilbert space represents a valid density matrix of the system.
	
	We note that the operators $\hat{O}$ can be classified into different groups according to the corresponding factors $(\delta_o,\gamma_o)$ defined in Eq.(\ref{eq:matrixcond1}) and (\ref{eq:matrixcond2}). Therefore, due to the two-valued properties of $\delta_o$ and $\gamma_o$, all the operators $\hat{O}$ can be divided into four categories ($\hat{A},\hat{B},\hat{C},\hat{D}$) and are listed as follows:
	\begin{enumerate}
		\item{ $\hat{A}:(\delta_o,\gamma_o)=(1,1)$
			\bea
			|\hat{A}+\zeta\hat{A}\mathcal{M}\rangle\rangle&=&|\hat{A}(1+\zeta\mathcal{M})\rangle\rangle =|\hat{A}_{+}\rangle\rangle[(1+\zeta)|1\rangle\rangle-(1-\zeta)|0\rangle\rangle];
			\eea
		}
		\item{ $\hat{B}:(\delta_o,\gamma_o)=(-1,1)$
			\bea
			|-\zeta\hat{B}+\hat{B}\mathcal{M}\rangle\rangle&=&|\hat{B}\mathcal{M}(1-\zeta\mathcal{M}\rangle\rangle=|\hat{B}_{+}\rangle\rangle[-(1+\zeta)|1\rangle\rangle+(1-\zeta)|0\rangle\rangle];
			\eea
		}
		\item{ $\hat{C}:(\delta_o,\gamma_o)=(1,-1)$
			\bea
			|\zeta\hat{C}+\hat{C}\mathcal{M}\rangle\rangle&=&|\hat{C}\mathcal{M}(1+\zeta\mathcal{M}\rangle\rangle =|\hat{C}_{+}\rangle\rangle[(1+\zeta)|1\rangle\rangle-(1-\zeta)|0\rangle\rangle];
			\eea
		}
		\item{ $\hat{D}:(\delta_o,\gamma_o)=(-1,-1)$
			\bea
			|\hat{D}-\zeta\hat{D}\mathcal{M}\rangle\rangle&=&|\hat{D}(1-\zeta\mathcal{M}))\rangle\rangle =|\hat{D}_{+}\rangle\rangle[(1-\zeta)|0\rangle\rangle-(1+\zeta)|1\rangle\rangle].
			\eea
		}
	\end{enumerate}
	We also note that to ensure the Hermiticity of $\rho$, these operators $\{\hat{A},\hat{B},\hat{C},\hat{D}\}$ also should be chosen to commute with $\mathcal{M}$.
	Therefore, the most general form of $|\rho_+\rangle\rangle$ reads
	\bea
	|\rho_+\rangle\rangle &=& |I_+\rangle\rangle+\sum_i a_{i}|\hat{A}_{i,+}\rangle\rangle+\sum_j b_{j}|\hat{B}_{j,+}\rangle\rangle +\sum_k  c_{k}|\hat{C}_{k,+}\rangle\rangle+\sum_l d_{l}|\hat{D}_{l,+}\rangle\rangle,
	\eea
	where all the coefficients $a_i$, $b_j$, $c_k$, and $d_l$ are real.
	The corresponding density matrix can be obtained accordingly and reads
	\bea
	\rho &=& \frac{1}{2^N} \bigg[ (I+\sum_{i}a_{i}\hat{A}+\sum_{k}c_{k}\hat{C_{k}}\mathcal{M})(I+\zeta\mathcal{M}) -(\sum_{j}b_{j}\hat{B_{j}}\mathcal{M}+\sum_{l}d_{l}\hat{D}_{l})(I-\zeta\mathcal{M})\bigg],  \label{eq:productmatrix}
	\eea
	where both the coefficients $(a_i,b_j,c_k,d_l)$ and the operators $(\hat{A}_i,\hat{B}_j,\hat{C}_k,\hat{D}_l)$ are carefully chosen so that $\rho$ is positive definite. In the special case with $b_j=d_l=0$ for all $j$ and $l$, the positivity of $\rho$ is reduced to find $(a_i,c_k)$ and $(\hat{A}_i,\hat{C}_k)$ such that $(I+\sum_i a_{i}\hat{A}_{i}+\sum_k c_{k}\hat{C}_{k}\mathcal{M})$ is positive defined, as shown in the main text.
	For general case, to ensure the positivity of $\rho$, a sufficient condition can be chosen such that both $(I+\sum_i a_{i}\hat{A}_{i}+\sum_k c_{k}\hat{C}_{k}\mathcal{M})$ and $-(\sum_{j}b_{j}\hat{B_{j}}\mathcal{M}+\sum_{l}d_{l}\hat{D}_{l})$ are positive operators.
	
	We also note that any Hermitian observable operator $\hat{X}$ which maps to a product vector in Liouville-Fock space can also be constructed following the above discussions. For instance, all operators defined in Eq.(\ref{eq:productmatrix}) are product in Liouville-Fock space.
	If we choose the two operators $\hat{X}_1$ and $\hat{X}_2$ as  $(\hat{X}_1,\hat{X}_2)=(\hat{O},\hat{O}\mathcal{M})$, then the ratio $\langle \hat{X}_1 \rangle /\langle \hat{X}_2 \rangle $ can be simplified as
	\bea
	\frac{\langle X_{1}\rangle}{\langle X_{2}\rangle}=\frac{\langle\langle X_{1}|\rho\rangle\rangle}{\langle\langle X_{2}|\rho\rangle\rangle}&=&\frac{\delta_{0}+\gamma_{0}+\delta_{0}\zeta(\delta_{0}-\gamma_{0})}{\delta_{0}-\gamma_{0}+\delta_{0}\zeta(\delta_{0}+\gamma_{0})}=\delta_{0}\zeta^{-\delta_{0}},
	\eea
	where in the last step, we have used the two-valued properties of $\delta_o$ and $\gamma_o$.
	Therefore $\langle \hat{X}_1 \rangle/\langle \hat{X}_2 \rangle$ is time-independent during the evolution for the initial product state if the edge mode is decoupled form all the bulk modes in Liouville-Fock space. This can be used to clarify the existence of LMEMs in this dissipative system.
	
	The edge modes are topologically protected by the internal symmetry of the system. The influences of perturbations on the system can be characterised by  introducing additional interaction  $H'$ to the Hamiltonian $H$, or new dissipator  $L'$ into the Lindblad equation.
	The edge modes are decoupled from the bulk modes as long as the corresponding Lindbladians in Liouville-Fock space are commuted with  $\kappa_1$ and $\kappa_{4N}$, namely, $[X_{H'},\kappa_{1}]=[X_{H'},\kappa_{4N}]=0$.  Since
	\bea
	\kappa_{1}|\rho\rangle\rangle &\rightarrow&-\sigma_{1}^{x}\mathcal{M}\rho\mathcal{M}\sigma_{1}^{x}, \nn \\
	\kappa_{4N}|\rho\rangle\rangle &\rightarrow& i \sigma_{N}^{x}\rho \mathcal{M} \sigma_{N}^{x}, \\
	X_{H'}|\rho\rangle\rangle &\rightarrow& [H',\rho], \nn \\
	X_{L'}|\rho\rangle\rangle &\rightarrow& 2L'^{\dag}\rho L' - L'L'^{\dag}\rho - \rho L'L'^{\dag}. \nn
	\eea
	Using the spin language, we can rewrite $[X_{H'},\kappa_{1}]|\rho\rangle\rangle=0$ as
	\bea
	[\mathcal{M}\sigma_{1}^{x}H'\sigma_{1}^{x}\mathcal{M}-H',\rho]=0,
	\eea
	which is valid for any given density matrix $\rho$. This leads to the following constraints for $H'$ as
	\bea
	[H',\sigma_{N}^{x}]=[H',\sigma_{1}^{x}]=[H',\mathcal{M}]=0.
	\eea
	Similar discussions also hold for additional dissipator $L'$ by noticing $[X_{L'},\kappa_{1}]|\rho\rangle\rangle=0$, and we have
	\bea
	2(\mathcal{M}\sigma_{1}^{x}L'^{\dagger}\sigma_{1}^{x}\mathcal{M}\rho \mathcal{M}\sigma_{1}^{x}L'\sigma_{1}^{x}\mathcal{M}-L'^{\dagger}\rho L') &-& (\mathcal{M}\sigma_{1}^{x}L'L'^{\dagger}\sigma_{1}^{x}\mathcal{M}
	-L'L'^{\dagger})\rho \nn \\&-& \rho(\mathcal{M}\sigma_{1}^{x}L'L'^{\dagger}\sigma_{1}^{x}\mathcal{M}-L'L'^{\dagger})=0.
	\eea 	
	To ensure that the above identity holds for any density matrix $\rho$, we have
	\bea
	[L',\sigma_{N}^{x}]=[L',\sigma_{1}^{x}]=[L',\mathcal{M}]=0.
	\eea

	In the main text, the existence of LMEMs is verified for different initial states and observables. Both of them can be re-expressed as bulk-edge product vectors in Liouville-Fock space. Specifically, for  $N=8$ and
	$\rho_0=[I+\sum_{j=2}^{N-1}0.2(\sigma_{j}^{x}\sigma_{j+1}^{x}+\sigma_{j}^{z})](I+0.5\prod_{j=1}^{N}\sigma_{j}^{z})/2^{N}$, all the corresponding operators $M_j=\sigma_{j}^{x}\sigma_{j+1}^{x}+\sigma_{j}^{z}$ ($2\le j\le N-1$) satisfy $(\delta_M,\gamma_M)=(1,1)$ and belongs to $\hat{A}$-category discussed above.  The relevant vector of $\rho_0$  is product in Liouville-Fock space and reads
	\bea
	|\rho_{0}\rangle\rangle=(|I_{+}\rangle\rangle+0.2\sum_{j=2}^{N-1}|M_{j,+}\rangle\rangle)(1.5|1\rangle\rangle-0.5|0\rangle\rangle)/2^{N},
	\eea
	with $\zeta=0.5$ ,  $M_{j}=\sigma_{j}^{x}\sigma_{j+1}^{x}+\sigma_{j}^{z}$.
	Similarly, using the following maps
	\bea
	\sigma_{j}^{z}  &\Longrightarrow&  |\sigma_{j,+}^{z}\rangle\rangle(|1\rangle\rangle-|0\rangle\rangle), (j\neq 1,N) \\
	\sigma_{j}^{x}\sigma_{j+1}^{x}  &\Longrightarrow&  |(\sigma_{j}^{x}\sigma_{j+1}^{x})_{+}\rangle\rangle(|1\rangle\rangle-|0\rangle\rangle).
	\eea
	We can find that the relevant state vectors in Liouville-Fock space for the observables $X_{1}=\sum_{j=2}^{N-1}(\sigma_{j}^{x}\sigma_{j+1}^{x}+\sigma_{j}^{z})$ and $X_2=X_1\mathcal{M} $ can be written as
	\bea
	|X_1\rangle\rangle =|M_+\rangle\rangle(|1\rangle\rangle-|0\rangle\rangle), \hspace{1cm}
	|X_2\rangle\rangle =|M_+\rangle\rangle(|1\rangle\rangle+|0\rangle\rangle),
	\eea
	where
	\bea
	|M_+\rangle\rangle = |\sum_{j=2}^{N-1}(\sigma_{j}^{x}\sigma_{j+1}^{x}+\sigma_{j}^{z})_{+}\rangle\rangle.
	\eea

	\section{ \label{E} Correlation $\langle\langle i\kappa_{1}\kappa_{4N}\rangle\rangle$ in Liouville-Fock space and the Purity of $\rho$}
	
	Since any density matrix $\rho$ can be expanded using pairs of Hermitian operators $\{\hat{O}_j,\hat{O}_j\mathcal{M}\}$, the corresponding state vector in Liouville-Fock space can always be written as
	\bea
	|\rho\rangle\rangle = \sum_j r_j |\rho_j\rangle \rangle = \sum_j r_j|\hat{O}_{j,+}\rangle\rangle (a_j |1\rangle\rangle + b_j |0\rangle\rangle).
	\eea
	Therefore the occupation number  $\langle\langle\rho(t)|c_{e}^{\dagger}c_{e}|\rho(t)\rangle\rangle$  of the edge mode for the given vector $|\rho(t)\rangle\rangle$ in Liouville-Fock space reads
	\bea
	\langle\langle\rho(t)|c_{e}^{\dagger}c_{e}|\rho(t)\rangle\rangle &=& \sum_{ij}r_{i}^{*}r_{j}\langle\langle\rho_{i}|c_{e}^{\dagger}c_{e}|\rho_{j}\rangle\rangle = \sum_{ij}r_{i}^{*}r_{j}a_{i}^{*}a_{j}\langle\langle\hat{O}_{i,+}|\hat{O}_{j,+}\rangle\rangle.
	\eea
	Using the relations $\langle\langle\hat{O}_{i,+}|\hat{O}_{j,+}\rangle\rangle =  2^{N-1}\delta_{ij}$, and $i\kappa_{1}\kappa_{4N} = 2c_{e}^{\dagger}c_{e}-1$,
	we immediately obtain
	\bea
	\langle\langle\rho(t)|i\kappa_{1}\kappa_{4N}|\rho(t)\rangle\rangle=2^{N-1}\sum_{i}r_{i}^{2}(a_{i}^{2}-b_{i}^{2}).
	\eea
	Meanwhile, the purity $\tr(\rho^2)$ of the density matrix $\rho$ can be re-expressed in Liouville-Fock space as
	\bea
	\langle\langle\rho(t)|\rho(t)\rangle\rangle&=&\sum_{ij}r_{i}^{*}r_{j}\langle\langle\rho_{i}|\rho_{j}\rangle\rangle=2^{N-1}\sum_{i}r_{i}^{2}(a_{i}^{2}+b_{i}^{2}).
	\eea
	This means that for bulk-edge product state in Liouville-Fock space with $(a_j,b_j)=(a,b)$ for all $j$, the correlation $\langle\langle i\kappa_{1}\kappa_{4N}\rangle\rangle=\langle\langle\rho(t)|i\kappa_{1}\kappa_{4N}|\rho(t)\rangle\rangle$ is directly linked with $\tr(\rho^2)$, and satisfies
	\bea
	\langle\langle i\kappa_{1}\kappa_{4N}\rangle\rangle=\frac{a^{2}-b^{2}}{a^{2}+b^{2}}\tr(\rho^{2}).
	\eea
	
	For the initial state discussed in the main text
	\bea
	\rho_0&=&[I+0.3\prod_{j=1}^{N}\sigma_{j}^{z}(\sigma_1^z+\sigma_1^y\sigma_2^x+\sigma_1^y\sigma_3^x +\sigma_2^z\sigma_2^x\sigma_3^x)](I+0.4\prod_{j=1}^{N}\sigma_{j}^{z})/2^{N},
	\eea
	the corresponding vector in Liouville-Fock space can be written as
	\bea
	|\rho_0\rangle\rangle = |\rho_{0,+}\rangle\rangle \otimes [(1+\zeta)|1\rangle\rangle-(1-\zeta)|0\rangle\rangle]/2^N
	\eea
	with $\zeta=0.4$ and
	\bea
	|\rho_{0,+}\rangle\rangle &=&
	|I_{+}\rangle\rangle+
	0.3(|(\sigma_1^z)_+\rangle\rangle+|(\sigma_1^y\sigma_2^x)_+\rangle\rangle +|(\sigma_1^y\sigma_3^x)_+\rangle\rangle+|(\sigma_2^z\sigma_2^x\sigma_3^x)_+\rangle\rangle).
	\eea
	This state has nonzero components in subspaces defined by $\{p_j=1 (\forall j)\}$ and $\{p_{j\ne1}=1(\forall j),p_{1}=-1\}$.
	
	For larger $\gamma$, the excited states for eigenvalue $\lambda$ with the minimum $|\Im(\lambda)|>0$ are degenerate in the subspace $\{p_{j\ne1}=1(\forall j),p_{1}=-1\}$ and read
	\bea
	\rho_{1}^{0}&=&(\alpha\sigma_{1}^{z}+\sigma_{1}^{y}\sigma_{2}^{x})\prod_{j=1}^{N}\sigma_{j}^{z},  \\
	\rho_{1}^{1}&=&\alpha\sigma_{1}^{y}\sigma_{2}^{x}+\sigma_{1}^{z}
	\eea
	with $\alpha = (i\gamma\pm\sqrt{\gamma^{2}-J^{2}})/J$. The above excited states and the stable states can then be viewed as the combinations of following operators
	\bea
	M =\{I,\prod_{i}\sigma_{i}^{z},\sigma_{1}^{y}\sigma_{2}^{x},\sigma_{1}^{z}\} \cup \{I,\prod_{i}\sigma_{i}^{z},\sigma_{1}^{y}\sigma_{2}^{x},\sigma_{1}^{z}\}\prod_{i}\sigma_{i}^{z}\nn
	\eea
	The purity $\tr(\rho^{2})$ can be approximated as
	\bea
	\langle\langle\rho|\rho\rangle\rangle=\tr(\rho^{2}) \simeq \frac{1}{2^{N}}\sum_{O_i \in M}\langle O_{i}\rangle^{2}.
	\eea
	For the given initial state $|\rho_0\rangle\rangle$, since the following relations hold
	\bea
	\langle\sigma_{1}^{y}\sigma_{2}^{x}\prod_{i}\sigma_{i}^{z}\rangle =\zeta\langle\sigma_{1}^{y}\sigma_{2}^{x}\rangle, \hspace{0.5cm}
	\langle\sigma_{1}^{z}\prod_{i}\sigma_{i}^{z}\rangle =\zeta\langle\sigma_{1}^{z}\rangle, \hspace{0.5cm}
	\langle\prod_{i}\sigma_{i}^{z}\rangle=\zeta,
	\eea
	we finally have
	\bea
	\langle\langle\rho|i\kappa_{1}\kappa_{2N}|\rho\rangle\rangle &\simeq&\frac{1}{2^{N-1}}\zeta \Big(1+\langle\sigma_{1}^{z}\rangle^{2}+\langle\sigma_{1}^{y}\sigma_{2}^{x}\rangle^{2} \Big).
	\eea	

\end{widetext}
	
	\bibliography{cite2023}
	
\end{document}